\documentclass[]{aastex631}
\usepackage{multirow}
\usepackage{xltabular}
\usepackage{booktabs}
\shorttitle{Optical and $\gamma-$ray variability of PKS~1222+216}
\shortauthors{Ezhikode et al.}

\begin{document}

\title{Long-term optical and $\gamma-$ray variability of the blazar PKS~1222+216}

\email{E-mail: savithri.ezhikode@christuniversity.in}

\author[0000-0003-1795-3281]{Savithri H. Ezhikode}
\affiliation{Inter-University Centre for Astronomy and Astrophysics, Post Bag 4, Ganeshkhind, Pune 411007, India}
\affiliation{Department of Physics and Electronics, CHRIST (Deemed to be University), Hosur Road, Bengaluru 560029, India}

\author[0000-0002-5656-2657]{Amit Shukla}
\affiliation{Department of Astronomy, Astrophysics and Space Engineering, Indian Institute of Technology Indore, Khandwa Road, Simrol, Indore, 453552 India}

\author[0000-0003-1589-2075]{Gulab C. Dewangan}
\affiliation{Inter-University Centre for Astronomy and Astrophysics, Post Bag 4, Ganeshkhind, Pune 411007, India}

\author[0000-0003-3188-1501]{Pramod K. Pawar}
\affiliation{Inter-University Centre for Astronomy and Astrophysics, Post Bag 4, Ganeshkhind, Pune 411007, India}

\author[0000-0001-5507-7660]{Sushmita Agarwal}
\affiliation{Department of Astronomy, Astrophysics and Space Engineering, Indian Institute of Technology Indore, Khandwa Road, Simrol, Indore, 453552 India}

\author[0000-0002-7254-191X]{Blesson Mathew}
\affiliation{Department of Physics and Electronics, CHRIST (Deemed to be University), Hosur Road, Bengaluru 560029, India}

\author[0000-0002-6096-3330]{Akhil Krishna R.}
\affiliation{Department of Physics and Electronics, CHRIST (Deemed to be University), Hosur Road, Bengaluru 560029, India}




\begin{abstract}
The $\gamma-$ray emission from flat-spectrum radio quasars (FSRQs) is thought to be dominated by the inverse Compton scattering of the external sources of photon fields, e.g., accretion disk, broad-line region (BLR), and torus. FSRQs show strong optical emission lines and hence can be a useful probe of the variability in BLR output, which is the reprocessed disk emission. We study the connection between the optical continuum, H$\gamma$ line, and $\gamma-$ray emissions from the FSRQ PKS~1222+216, using long-term ($\sim$2011--2018) optical spectroscopic data from Steward Observatory and $\gamma-$ray observations from \textit{Fermi}-LAT. We measured the continuum ($F_{\rm{C,opt}}$) and H$\gamma$ ($F_{\rm{H\gamma}}$) fluxes by performing a systematic analysis of the 6029--6452~{\AA} optical spectra. We observed stronger variability in $F_{\rm{C,opt}}$ than $F_{\rm{H\gamma}}$, an inverse correlation between H$\gamma$ equivalent width and $F_{\rm{C,opt}}$, and a redder-when-brighter trend. Using discrete cross-correlation analysis, we found a positive correlation (DCF$\sim$0.5) between $F_{\rm{\gamma-ray>100MeV}}$ and $F_{\rm C,opt}$~(6024--6092{\AA}) light curves with time-lag consistent with zero at 2$\sigma$ level. We found no correlation between $F_{\rm{\gamma-ray>100MeV}}$ and $F_{\rm{H\gamma}}$ light curves, probably dismissing the disk contribution to the optical and $\gamma$-ray variability. The observed strong variability in the \textit{Fermi}-LAT flux and $F_{\rm{\gamma-ray>100MeV}}-F_{\rm{C,opt}}$ correlation could be due to the changes in the particle acceleration at various epochs. We derived the optical-to-$\gamma$-ray spectral energy distributions (SEDs) during the $\gamma$-ray flaring and quiescent epochs that show a dominant disk component with no variability. Our study suggests that the $\gamma$-ray emission zone is likely located at the edge of the BLR or in the radiation field of the torus.
\end{abstract}



\section{Introduction} \label{sec:intro}

Blazars are sources with highly collimated jets that point very close to the line of sight of the observer \citep{1995PASP..107..803U}. The radiation from blazars ranges from radio to high energy $\gamma-$rays, and their non-thermal emission is dominated by the output from the jet. Blazars are highly variable sources, and they show flaring activities at multiple wavebands that may be correlated or uncorrelated \citep[e.g.][]{2018MNRAS.480.5517L, 2019ApJ...880...32L, 2019ApJ...877...39M}.

The spectral energy distribution (SED) of blazars has a double-hump structure \citep{1996ASPC..110..391U, 1996ApJ...463..444S}. The low-energy hump typically peaks around infrared to X-ray regions and is attributed to synchrotron emission from electrons present in the jet. The high-energy component arising from the X-ray to $\gamma-$ray bands is believed to be produced either by Synchrotron Self Compton (SSC) process or External Comptonization (EC) process \citep[e.g.][]{1997ApJS..109..103D, 1999ApJ...515L..21B, 1997ApJ...490..116M, 1996MNRAS.280...67G,2002A&A...386..415A}. In the EC scenario, the external photon field could be the accretion disk, broad-line region (BLR), and dusty torus.

The relationship between the accretion disk and jet power is one of the major unresolved issues in active galactic nuclei (AGN). The thermal disk emission from AGN mostly emerges in the optical/UV bands. Since the optical/UV continuum in blazars is contributed by both the disk and jet emissions, it is difficult to get a direct measurement of the disk emission from the observed optical/UV luminosity. However, the radiation from the disk can photoionize the gas in broad-line and narrow-line regions, producing emission lines in the optical spectra. Then, any variability found in the optical emission lines could be an indicator of the changes in the accretion power. Therefore, the correlation between the variable BLR luminosity or flux with the $\gamma-$ray flux can provide an indirect way to find the connection between the accretion disk and jet emissions. Many studies have been carried out in a similar line to find this correlation in blazars \citep[e.g.,][]{2013NewA...18....1D, 2014MNRAS.445...81S}.

A broad classification of blazars is based on the optical spectra where objects with strong optical emission lines (rest-frame equivalent width, EW~$>$~5~{\AA}) are known as flat-spectrum radio quasars (FSRQs), and those with relatively featureless optical spectra are called BL Lacertae (BL Lac) objects \citep{1995PASP..107..803U}. FSRQs are more luminous, on average than BL Lacs, and the $\gamma-$ray emission from these objects is better explained by the EC model. We study an FSRQ PKS~1222+216 to probe the possible correlation between the optical and $\gamma-$ray emission components. The cross-correlation analysis of optical and $\gamma-$ray fluxes may indicate the possible location of $\gamma-$ray emission in the source.

PKS~1222+216 or 4C+21.35 (RA = 12h24m54.4s, DEC = +21d22m46s; redshift = 0.432) is one of the bright FSRQs in which very high energy emission was observed. It was first discovered in the $\gamma-$rays with Energetic Gamma Ray Experiment Telescope (EGRET) \citep{1999ApJS..123...79H}. Later, the source was also detected with \textit{Fermi}-LAT \citep{2010ApJ...715..429A} and MAGIC \citep{2011ApJ...730L...8A} when it was in a very high state. The MAGIC observations showed rapid variability in the VHE emission, and the GeV spectrum of the source was found to be hard \citep{2011ApJ...730L...8A}. The source was also observed in the radio \citep[e.g.][]{2016A&A...596A.106P}, infrared \citep[e.g.][]{2011ApJ...732..116M}, optical \citep[e.g.][]{2011arXiv1110.6040S, 2012MNRAS.424..393F}, UV and X-ray \citep[e.g.][]{2018ApJ...863...98P} wavebands. Epochs of flaring activities have been reported in the source. Detailed studies on these active and quiescent states have been carried out using \textit{Fermi}-LAT observations \citep{2019ApJ...877...39M, 2014ApJ...796...61K}. Quite a few studies are available on the multi-wavelength SED analysis of this source \citep{2011A&A...534A..86T, 2018ApJ...863...98P, 2021MNRAS.500.1127B, 2021MNRAS.504.1103R}.

PKS~1222+216 is one among the blazars monitored with \textit{Fermi}-LAT for a long time scale. The optical properties of the source were studied using the coordinated monitoring observations with  Steward Observatory (SO) \citep{2011arXiv1110.6040S, 2012MNRAS.424..393F}. The optical spectrophotometric study of the source during the first two years of the \textit{Fermi}-LAT observations did not reveal any connection between the variable optical and $\gamma-$ray flux \citep{2011arXiv1110.6040S}. They also did not find any variability in the broad optical emission lines. \cite{2012MNRAS.424..393F} found similar optical spectral properties in the source where they do not find any notable variations in the broad-line emissions. From the H$\beta$ line-width obtained from SO observations, they also found a black hole mass of $(5.0-11.4) \times 10^8 M_\odot$. The multi-wavelength SED studies (from radio to $\gamma-$ray) of various quiescent states suggest relevant disk-jet connection in the source \citep[e.g.,][]{2021MNRAS.504.1103R}. Though studies were carried out to understand the temporal correlation between the disk, BLR, and jet emission from the source at various flaring states, no attempts have been made to probe the correlations on longer time scales. This work investigates the relationship between these emission components via the cross-correlation studies of long-term optical continuum/line and $\gamma-$ray light curves of the source.

The paper is organized as follows. In $\S$\ref{sec:obs}, we give the details of the observations used in the work. In $\S$\ref{sec:analysis-opt}, we explain the analysis of optical spectra. $\S$\ref{sec:corr} gives the details of cross-correlation analysis and $\S$\ref{sec:sed} provides a description of the SED analysis. In $\S$\ref{sec:rslt} we summarize the results. $\S$\ref{sec:sum} provides a detailed summary and discussion of the study.

\section{Observations} \label{sec:obs}

PKS~1222+216 has been continuously monitored with \textit{Fermi}-LAT from the start of the mission. In support of these observations, an optical monitoring program has been carried out using the ground-based Steward observatory of the University of Arizona. In this work, to explore the multi-wavelength variability properties of the blazar PKS~1222+216, we use the optical and $\gamma-$ray data on the source from these coordinated observations.

\newpage
\subsection{Fermi-LAT}  \label{sec:obs-Fermi}

The \textit{Fermi}-LAT \citep{2009ApJ...697.1071A} is a pair-conversion $\gamma$-ray telescope. \textit{Fermi}-LAT scans the whole sky in 3 hours period due to its large field of view of 2.4 sr. The pass8 \textit{Fermi}-LAT $\gamma$-ray data\footnote{https://fermi.gsfc.nasa.gov/ssc/data/} ($>$100 MeV) of PKS~1222+216 have been analyzed using Science Tools version v10r0p5 by the \textit{Fermi}-LAT collaboration and user-contributed Enrico software (Sanchez \& Deil 2013). A circular region of 15$^\circ$ radius around the PKS~1222+216 was chosen for the analysis. A zenith angle cut of $90^{\circ}$, the GTMKTIME cut of DATA\_QUAL==1 $\&\&$ LAT\_CONFIG==1 together with the LAT event class == 128 and the LAT event type == 3 were used. Spectral analysis of the resulting data set was carried out by including {\it gll\_iem\_v06.fits} and the isotropic diffuse model {\it iso\_P8R2\_SOURCE\_V6\_v06.txt}. A log-parabola model was used to fit the energy spectrum of PKS 1222+216, and its flux and spectrum were determined using an unbinned gtlike algorithm based on the NewMinuit optimizer. The \textit{Fermi}-LAT light curve above 100 MeV were obtained through \emph{ScienceTools} and \emph{Fermipy}.

\subsection{Steward Observatory}  \label{sec:obs-SO}

\begin{figure*}[htp]
    \centering
    \includegraphics[scale=1.]{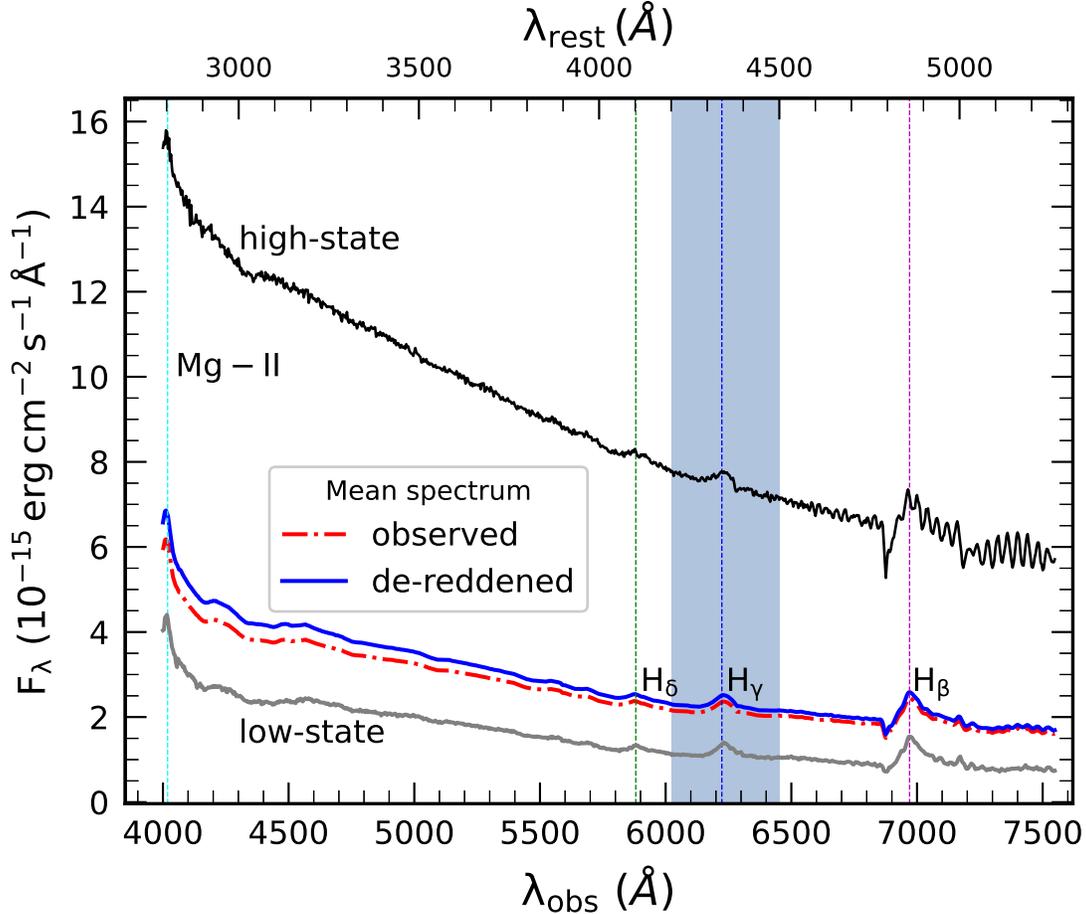}
    \caption{The optical spectra of PKS~1222+216 from SO SPOL observations. Observed and de-reddened \citep{1989ApJ...345..245C} mean spectra are plotted as dash-dot red and solid blue lines, respectively, along with the de-reddened spectra at high (black solid line) and low (solid gray line) flux states. The prominent emission lines, H$\beta$ ($\lambda_{\rm rest}=4861.3$~\AA), H$\gamma$ ($\lambda_{\rm rest}=4340.48$~\AA), H$\delta$ ($ \lambda_{\rm rest}=4101.75$~\AA) and \ion{Mg}{2} ($\lambda_{\rm rest}$=2795.50~{\AA}~\&~2802.7~{\AA}), in the observed frame are also identified in the plot. The shaded region shows the spectral region local to H$\gamma$ emission line that we used for the analysis.}
    \label{fig:opt_meanspec}
\end{figure*}

Steward Observatory uses the 2.3~m Bok Telescope on Kitt Peak and 1.54~m Kuiper Telescope on Mount Bigelow for the optical monitoring of blazars. The observations were performed with the high-throughput, moderate resolution spectropolarimeter SPOL \citep{2009arXiv0912.3621S}. The monitoring program provides the spectroscopic, photometric, and polarization data on each blazar observed. Here, we use the publicly available flux spectra of PKS~1222+216 from cycle 3 (C-3) to cycle 10 (C-10), corresponding to \textit{Fermi} cycles obtained from SPOL observations. These observations span roughly eight years, from 2011 to 2018. The optical spectra cover the wavelength range of 4000--7550~{\AA} with a dispersion of 4~\AA/pixel. The spectral resolution varies from 16~{\AA} to 24~{\AA} depending on the slit width (2"--12.7") used for the observation. The observations used in this work were taken with apertures of slit widths of 4.1", 5.1" and 7.6". The flux density spectra of the source corrected for airmass and instrument sensitivity are available at the website\footnote{\url{http://james.as.arizona.edu/~psmith/Fermi/DATA/Objects/pks1222.html}}. We select those observations where the flux spectra, averaged between 5400--5600~{\AA}, have been scaled to match the results from the synthetic V-band photometry on that night. The observed and de-reddened mean spectrum of these 329 observations, along with the spectra (de-reddened) at the low and high flux states are shown in Fig.~\ref{fig:opt_meanspec}. The strong emission lines observed, say H$\beta$ ($\lambda_{rest}~4861.3$~\AA), H$\gamma$ ($\lambda_{rest}=4340.48$~\AA), H$\delta$ ($\lambda_{rest}$=4101.75~{\AA}) and \ion{Mg}{2} ($\lambda_{rest}$=2795.50~{\AA} \& 2802.7~{\AA}), are also identified in the same figure.

\section{Optical Spectral Analysis}  \label{sec:analysis-opt}

\begin{figure*}[htp]
    \centering
    \includegraphics[scale=0.2]{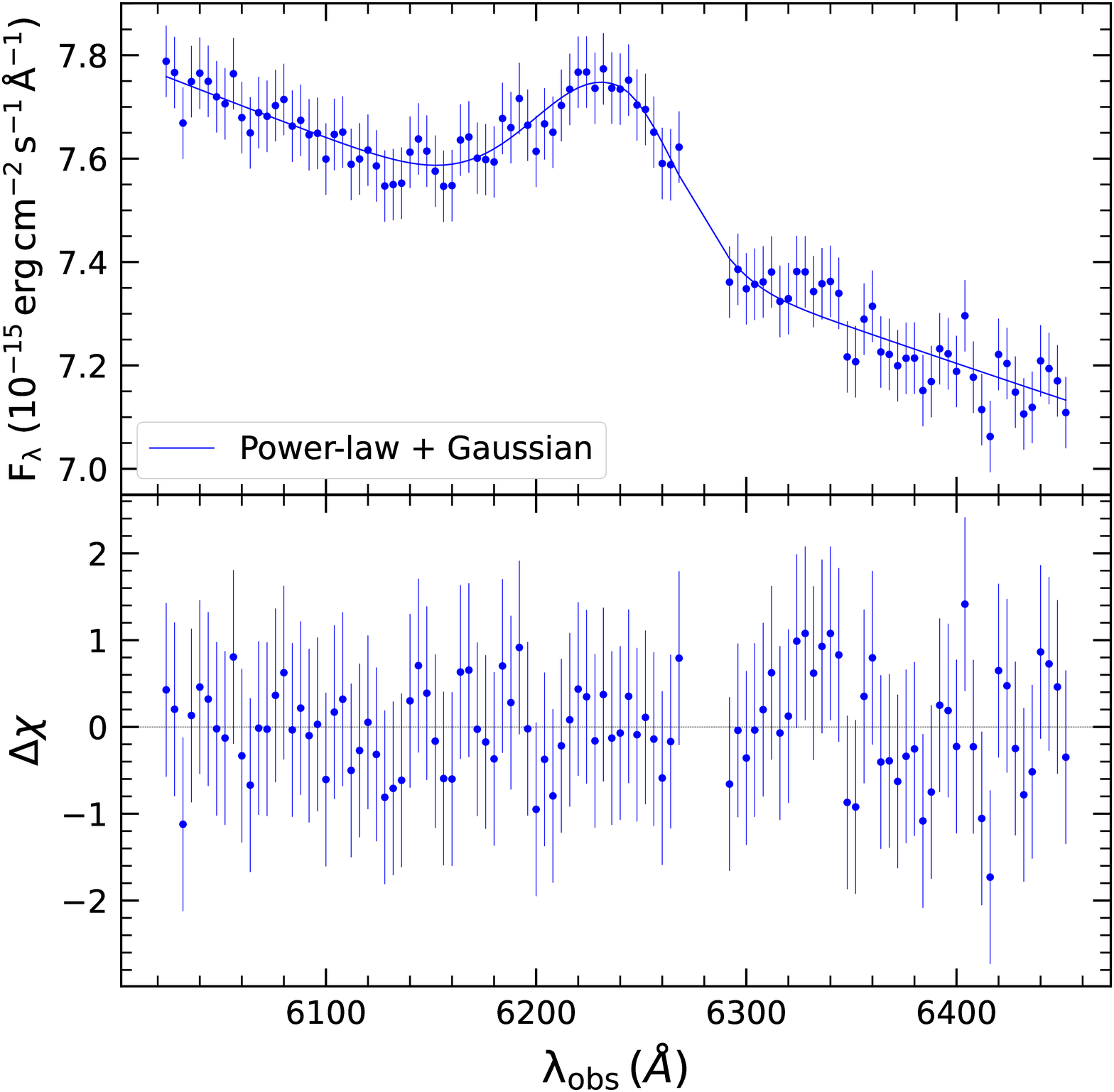}
    \includegraphics[scale=0.2]{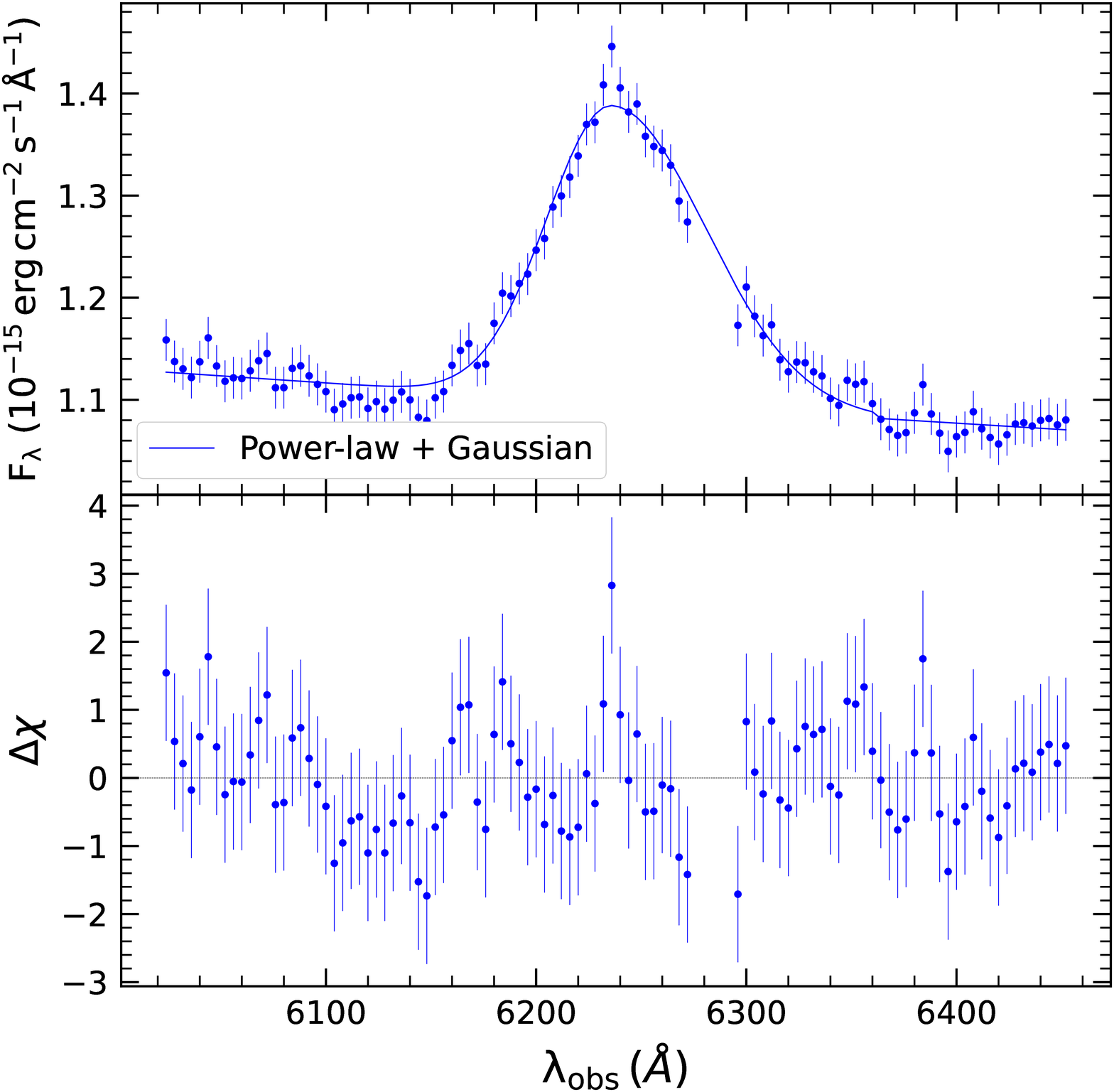}
    \caption{Spectral fitting plots of local continuum and H$\gamma$ line emission with \textit{powerlaw} and \textit{Gaussian} (after ignoring the region of atmospheric absorption feature) at high (left) and low (right) optical flux states. The upper panel in each plot shows the de-reddened spectrum and the best-fit model. The lower panels show the residuals associated with the spectral fits.}
    \label{fig:opt_spec_fit}
\end{figure*}

We derive the optical continuum and emission-line flux of the FSRQ PKS~1222+216. Since the optical continuum in blazars is contaminated by the jet emission, we need a different quantity to probe the disk emission. The optical emission line components from BLR and NLR are produced by the photoionization of the gaseous clouds in these regions by illuminating disk radiation. Hence, the variable broad optical emission lines can be used to track the changes in disk emission in FSRQs. To achieve this goal, we obtained the optical spectroscopic monitoring data on PKS~1222+216 in the observed wavelength range of 4000--7550~\AA, as mentioned in the previous section.

Since the available spectra are not corrected for Galactic extinction and reddening, we applied the correction method provided by \cite{1989ApJ...345..245C} with R$_V$=3.1 and A$_V$=0.077 \citep{1998ApJ...500..525S}. The errors on the spectra were calculated from the standard deviation of the flux density values in the wavelength range where there are no significant line features. The wavelength range higher than 7000~{\AA} is affected by the fringing of thinned CCD. Also, there could be atmospheric absorption features due to water vapor around 7200--7300~{\AA}. Hence, the spectral region above 7000~{\AA} is hard to analyze. Though H$\beta$ ($\lambda_{\rm obs} \sim 6970.3$~{\AA}) appears to be the strongest among the emission lines in the spectrum, the O$_2$ B-band ($\lambda_{\rm obs} \sim 6884$~\AA) absorption hinders the measurement of H$\beta$ line properties. Therefore, we chose the broad H$\gamma$ emission line to explore the variable emission from the BLR region. The continuum emission was determined from the local spectral region around the H$\gamma$ line. Here, we estimated the optical continuum as well as the emission line properties by modelling the flux spectra using Sherpa (Version 4.10.2) in Python (\url{https://doi.org/10.5281/zenodo.593753}). 

We used Sherpa optical models to analyze the local spectral region of $\sim6020-6450$~{\AA} (observer's frame) where the H$\gamma$ line is present. The continuum emission was modelled with \emph{powerlaw} component, and H$\gamma$ emission was modelled with a broad \emph{Gaussian} line model. The \emph{powerlaw} component has three parameters: the reference wavelength ($\lambda_{\rm ref}$ in {\AA}),  amplitude ($ampl$), and the index ($\alpha$). $\lambda_{\rm ref}$ was fixed at the average wavelength of the selected region, 6237~{\AA}, while the other two parameters were left free to vary. The parameters of \emph{Gaussian} component are the full-width half-maximum (\emph{FWHM}) in km/s, central wavelength (\emph{$\lambda_{\rm pos}$}) in {\AA}, \emph{Flux} which is the normalization of the \emph{Gaussian} and the skewness parameter \emph{skew}. All these parameters were thawed while fitting. Some of the observations showed atmospheric O$_2$ C-band absorption features around $6288$~\AA. Therefore, we removed this part of the spectrum during the analysis. Examples of the spectral fitting results are shown in Fig.~\ref{fig:opt_spec_fit}. 

The emission line flux ($F_{\rm H\gamma}$) values were directly obtained from the parameter \emph{Flux} of the best-fit \emph{Gaussian} line. The continuum fluxes ($F_{\rm C,opt}$) for two different regions, on either sides of the broad H$\gamma$ component, with $\lambda_{\rm obs} : 6024-6092$~{\AA} {\&} $~6380-6452$~{\AA} are calculated by integrating the power-law function defined in the model,
\begin{equation}
f(\lambda) = ampl~(\lambda/\lambda_{\rm ref})^{\alpha}.
\end{equation}

\begin{figure*}[htp]
    \centering
    \includegraphics[scale=0.42]{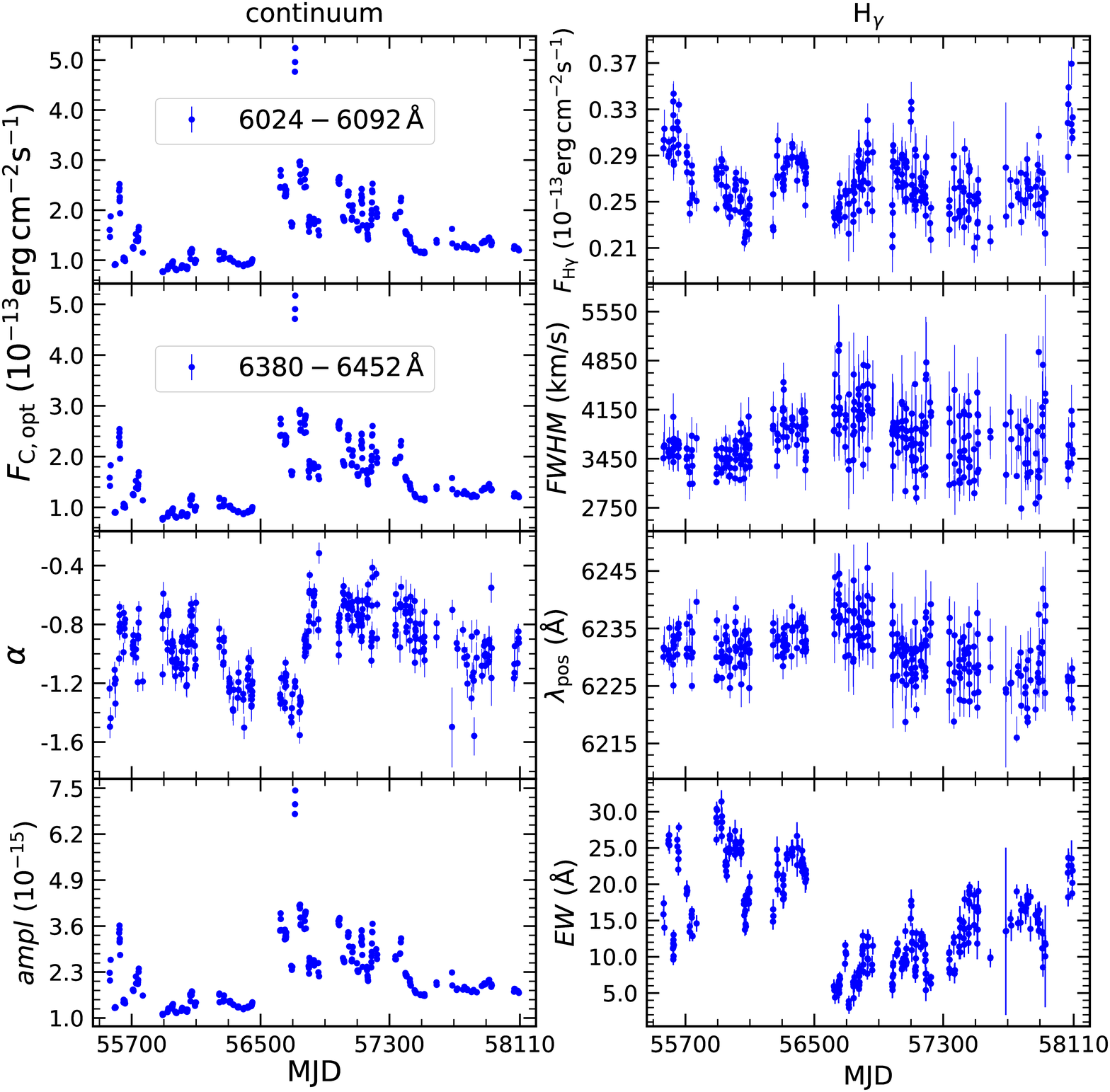}
    \caption{Best-fit parameters from the analysis of the optical spectral region ($6024-6452$~\AA) local to H$\gamma$ line using the models \textit{powerlaw} and \textit{Gaussian}. The left panels show the power-law properties such as the continuum flux ($F_{\rm C,opt}$) in the $6024-6092$~\AA and $6380-6452$~\AA wavebands (observer's frame), spectral index ($\alpha$) and amplitude (\textit{ampl}) of the model. The right panels give the emission line properties of H$\gamma$, say, line flux ($F_{H\gamma}$), \textit{FWHM}, best-fit central wavelength ($\lambda_{\rm pos}$) and the equivalent width (\textit{EW}) of the emission line.}
    \label{fig:opt_par}
\end{figure*}

\begin{figure*}[htp]
    \centering
    \includegraphics[scale=0.058]{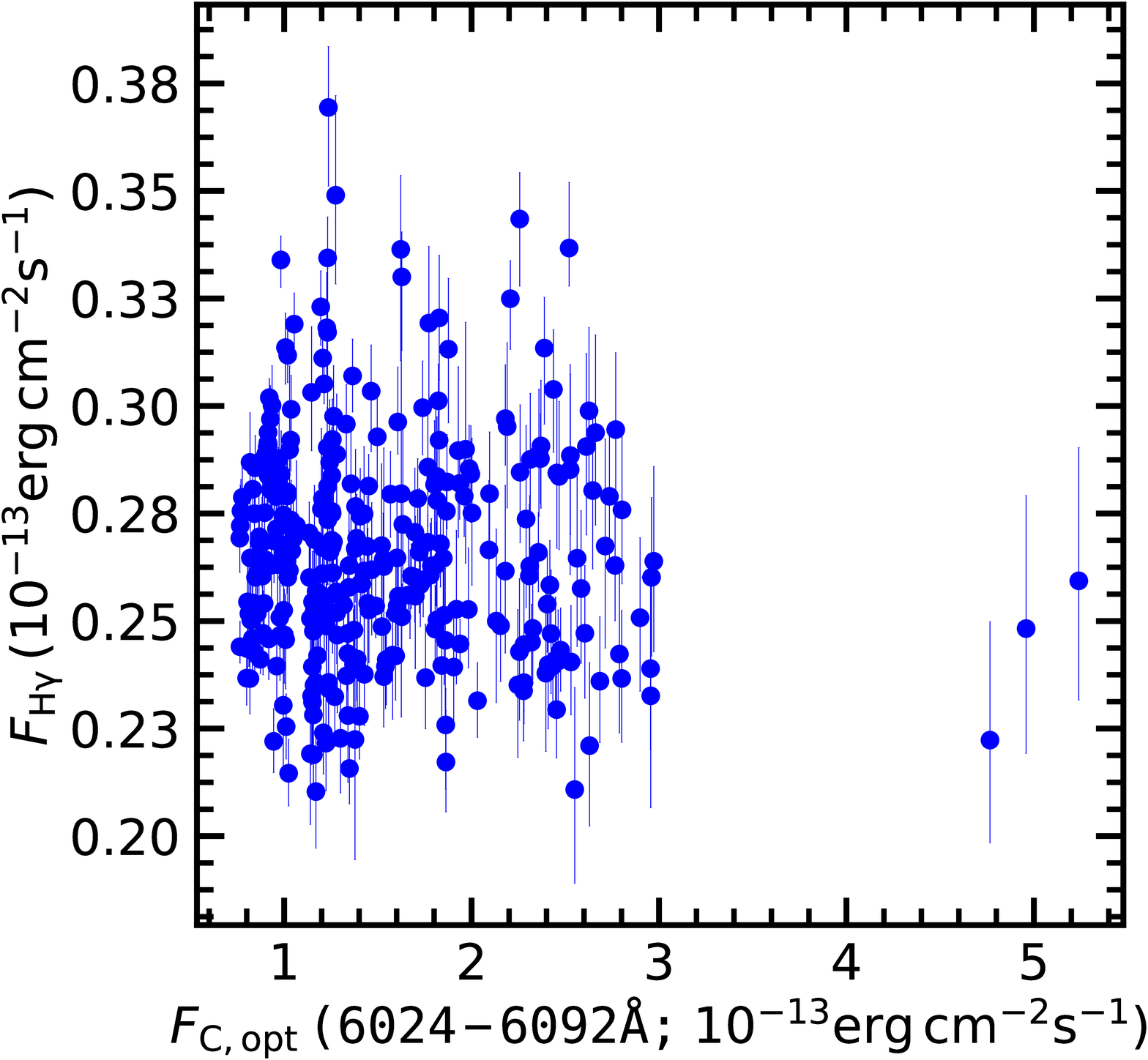}
    \includegraphics[scale=0.058]{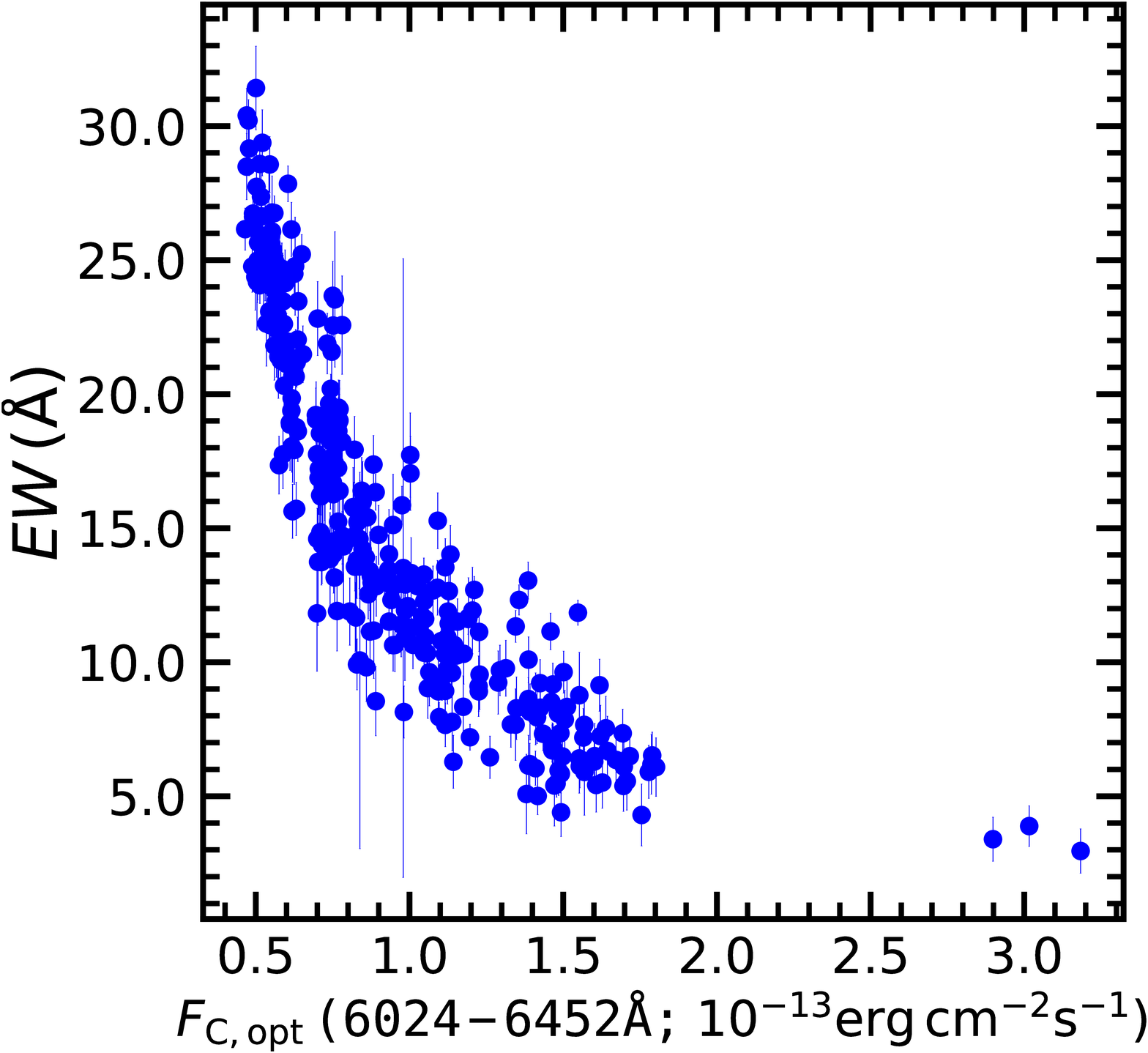}
    \includegraphics[scale=0.058]{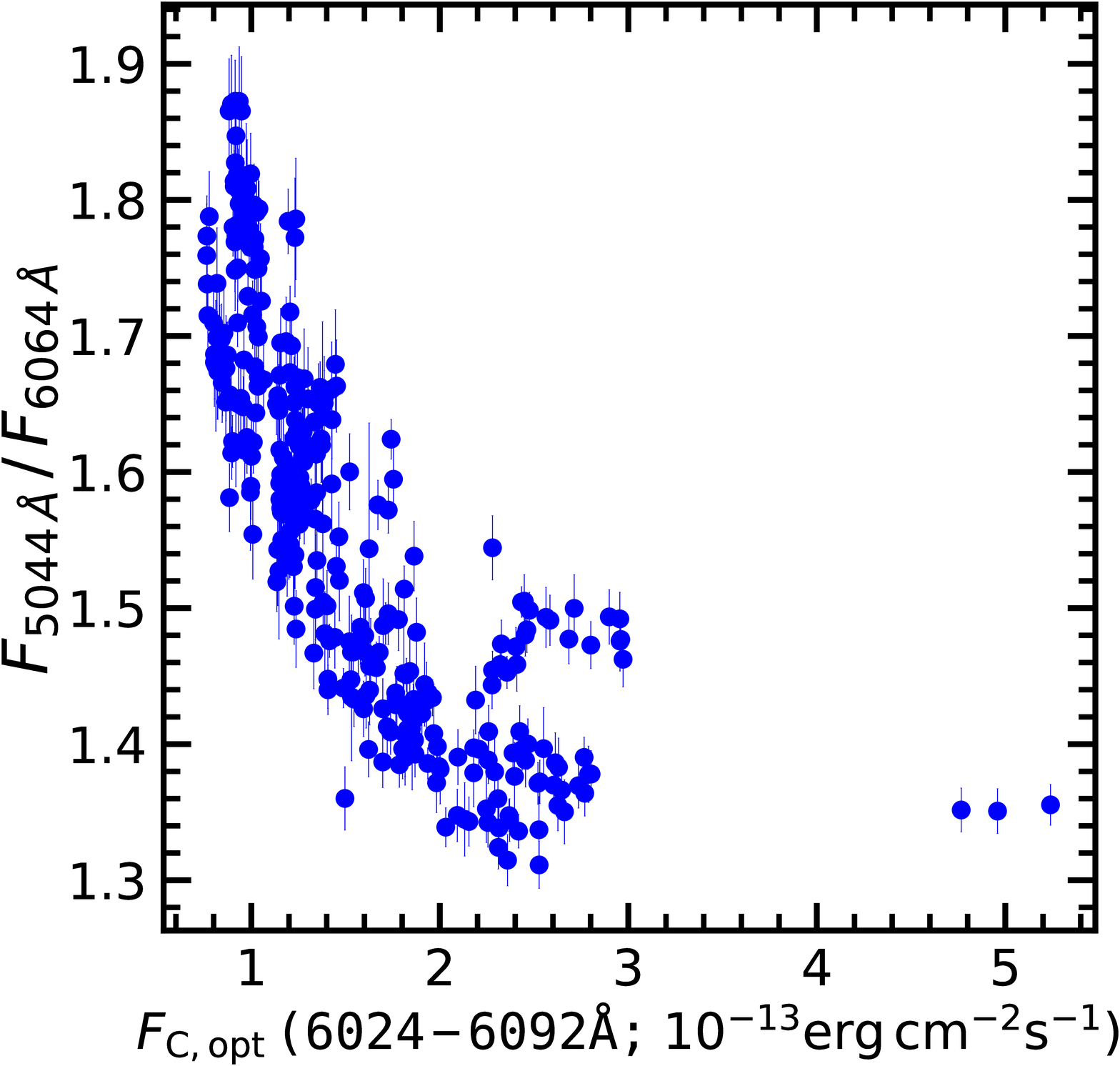}
    \caption{The variation of H$\gamma$ flux (left), equivalent width (middle), and the flux density ratio between 5044~\AA and 6064~{\AA} (right) with the narrow band optical continuum flux.}
    \label{fig:opt_cont_line_ew_Fratio}
\end{figure*}

The strength of an emission line can be estimated by measuring the equivalent width (\textit{EW}). We estimated the equivalent width of the H$\gamma$ line by integrating the quantity [($F_{\lambda}/F_{\lambda,cont}$)-1] over the spectral range local to H$\gamma$. Here, $F_{\lambda}$ and $F_{\lambda,cont}$ are respectively the total flux density and continuum flux density at each wavelength ($\lambda$) in the region. The H$\gamma$ equivalent widths, continuum fluxes, and the other parameters obtained from the spectral analysis of all the observations are shown in Fig.~\ref{fig:opt_par}. We also estimated the flux density ratio $F_{\rm 5044\AA}/F_{\rm 6064\AA}$ for wavelengths 5044~{\AA} and 6064~{\AA}. The dependence of the emission line flux, equivalent width and $F_{\rm 5044\AA}/F_{\rm 6064\AA}$ on the optical continuum flux are shown in Fig.~\ref{fig:opt_cont_line_ew_Fratio}. 

Please note that the entire optical spectral analysis has been done in the observed frame, and all the parameters we derived are in the same frame of reference.

\section{Temporal Correlations} \label{sec:corr}

\begin{figure*}[htp]
    \centering
    \includegraphics[scale=0.4]{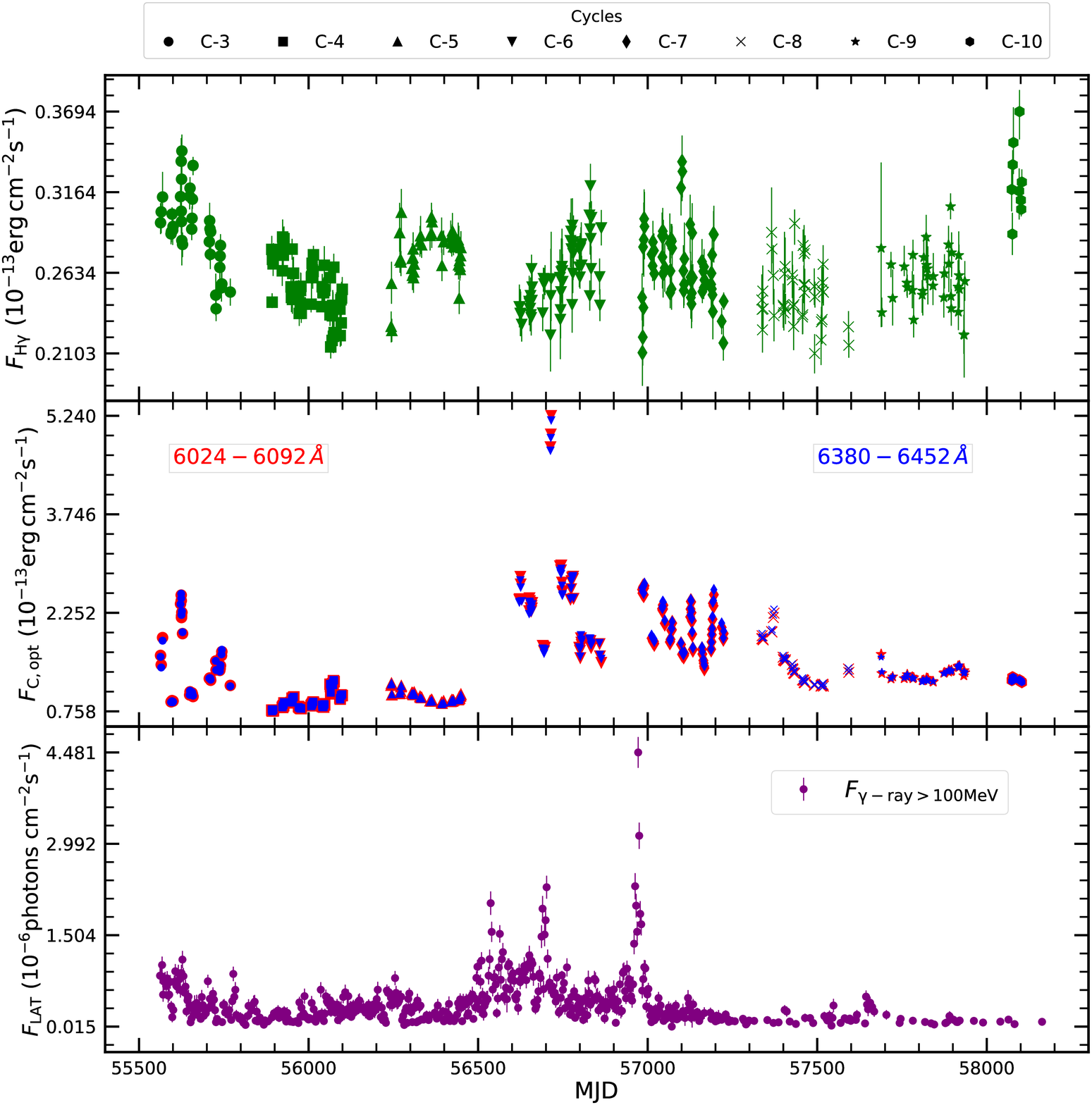}
    \caption{The light curves of H$\gamma$ emission line flux (upper), optical continuum fluxes for two wavelength regions (red: 6024-6092~{\AA} \& blue: 6380-6452~{\AA} ) (middle), and  $\gamma-$ray flux (\textit{Fermi}-LAT: 3$\sigma$ detection) (lower). Different markers for the optical light curves (upper and middle panels) represent various cycles used in the analysis. The error bars for the optical continuum flux values are smaller than the size of the markers, hence are not visible in the plot.}
    \label{fig:lc}
\end{figure*}

\begin{figure*}
     \centering
     \includegraphics[scale=0.2]{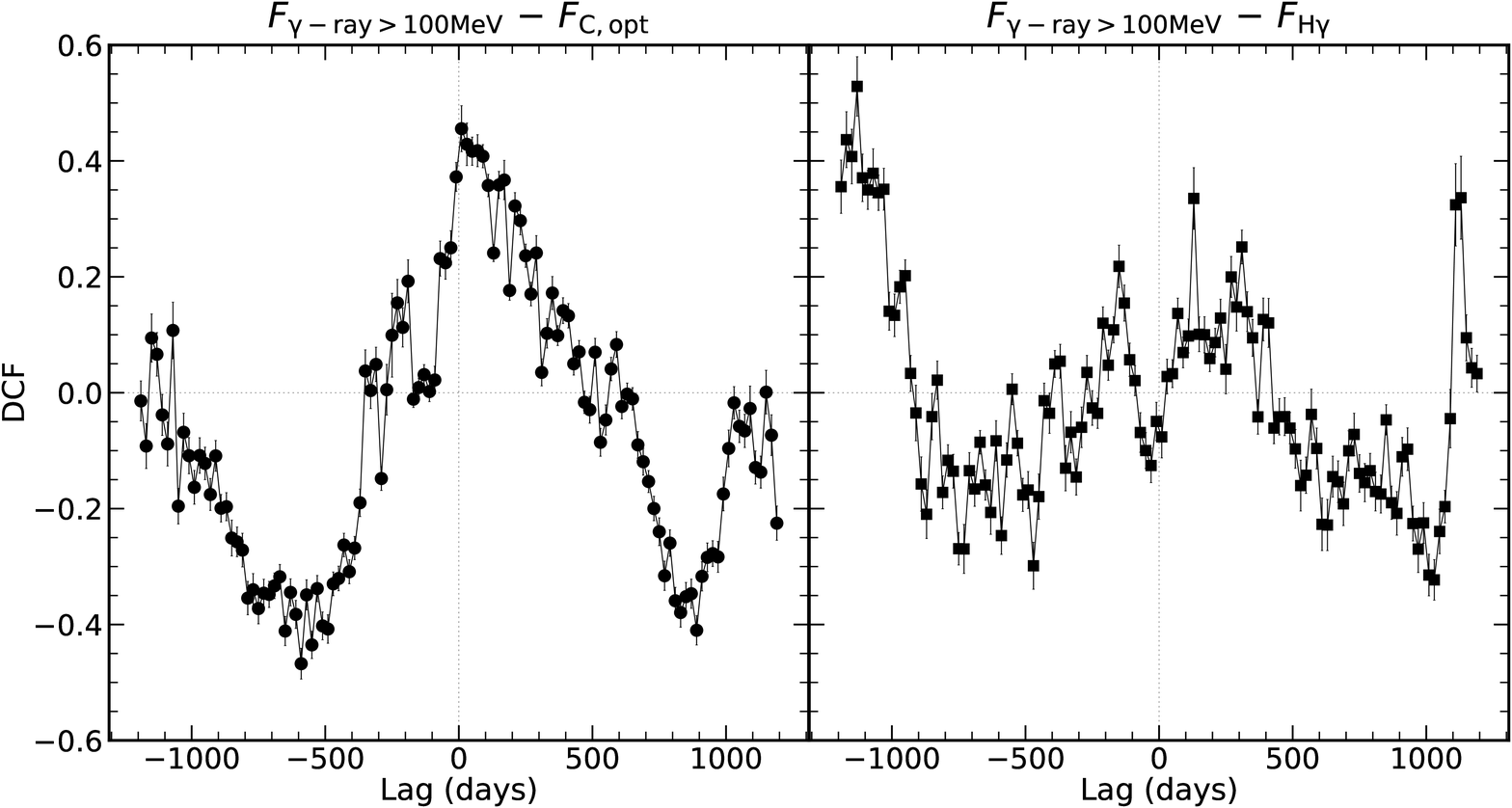}
     \caption{Cross-correlation plots obtained with PyDCF for a time range of $\pm$1200~days and lag bin size of 20~days for $\gamma-$ray ($F_{\rm {\gamma}-ray>100~MeV}$) and optical continuum ($F_{\rm C,opt}$~(6024--6092~{\AA})) light curves (left) {\&} $\gamma-$ray $F_{\rm {\gamma}-ray>100~MeV}$ and H$\gamma$ ($F_{\rm H{\gamma}}$) light curves (right).}
     \label{fig:dcf_fermi_cont_line}
\end{figure*}

The light curves of the optical continuum, H$\gamma$ line and $\gamma-$ray$_{>100~MeV}$ fluxes shown in Fig.~\ref{fig:lc} exhibit variability. The strength of variability of these emission components is discussed in the next section ($\S$\ref{sec:rslt}). To check the presence of any delayed response between the optical and $\gamma-$ray emissions from PKS~1222+216, we performed a cross-correlation analysis. Since our light curves have a lot of gaps, we used the discrete correlation method developed by \cite{1988ApJ...333..646E}. This method can perform the cross-correlation analysis of unevenly sampled time-series data without using any interpolation. We estimated the discrete correlation function (DCF) of the light curves using PyDCF\footnote{\url{https://github.com/astronomerdamo/pydcf}} \citep{2015MNRAS.453.3455R} method. The cross-correlation plots we obtained for $\gamma-$ray/optical continua and $\gamma-$ray/H$\gamma$ light curve pairs, for a lag range of $\pm 1200$~days and a lag bin of 20~days, are shown in Fig.~\ref{fig:dcf_fermi_cont_line}. No strong correlation is observed for $\gamma-$ray$_{>100~MeV}$/H$\gamma$ light curves for the time-lag range of $\pm1200$~days. However, the plot shows a moderately strong positive correlation between $\gamma-$ray$_{>100~MeV}$ and $F_{\rm C,opt}$~(6024--6092~{\AA}) light curves with a DCF value around 0.45. 

In order to find the confidence limits of the observed correlation between $\gamma-$ray and optical continuum fluxes, we followed the bootstrap method. We simulated 10000 random subsets of the original light curve pairs ($\gamma-$ray$_{>100~MeV}$ and $F_{\rm C,opt}$~(6024--6092~{\AA}) in our case). Each subset consists of a maximum of 80\% of the original data points. Then, using PyDCF, we estimated the cross-correlation of each ($\gamma-$ray/optical continuum) pair of the simulated light curves for a lag range of $\pm$1200 days and lag bin size of 20 days. The centroid DCF of the cross-correlation functions were determined from the average of the DCF values higher than 60\% of the maximum DCF. Then, we estimated the peak and the confidence intervals (68\%, 95\%, and 99\%) of the DCF from the distribution of the centroid values obtained with the simulations. To determine the significance of correlations, we estimated the 1$\sigma$, 2$\sigma$, and 3$\sigma$ confidence levels for the distributions of simulated DCF after subtracting the original DCF value for each lag. We found that the observed correlation is significant at 3$\sigma$ level over a broad range of lags.

\begin{figure*}[htp]
     \centering
     \includegraphics[scale=0.1]{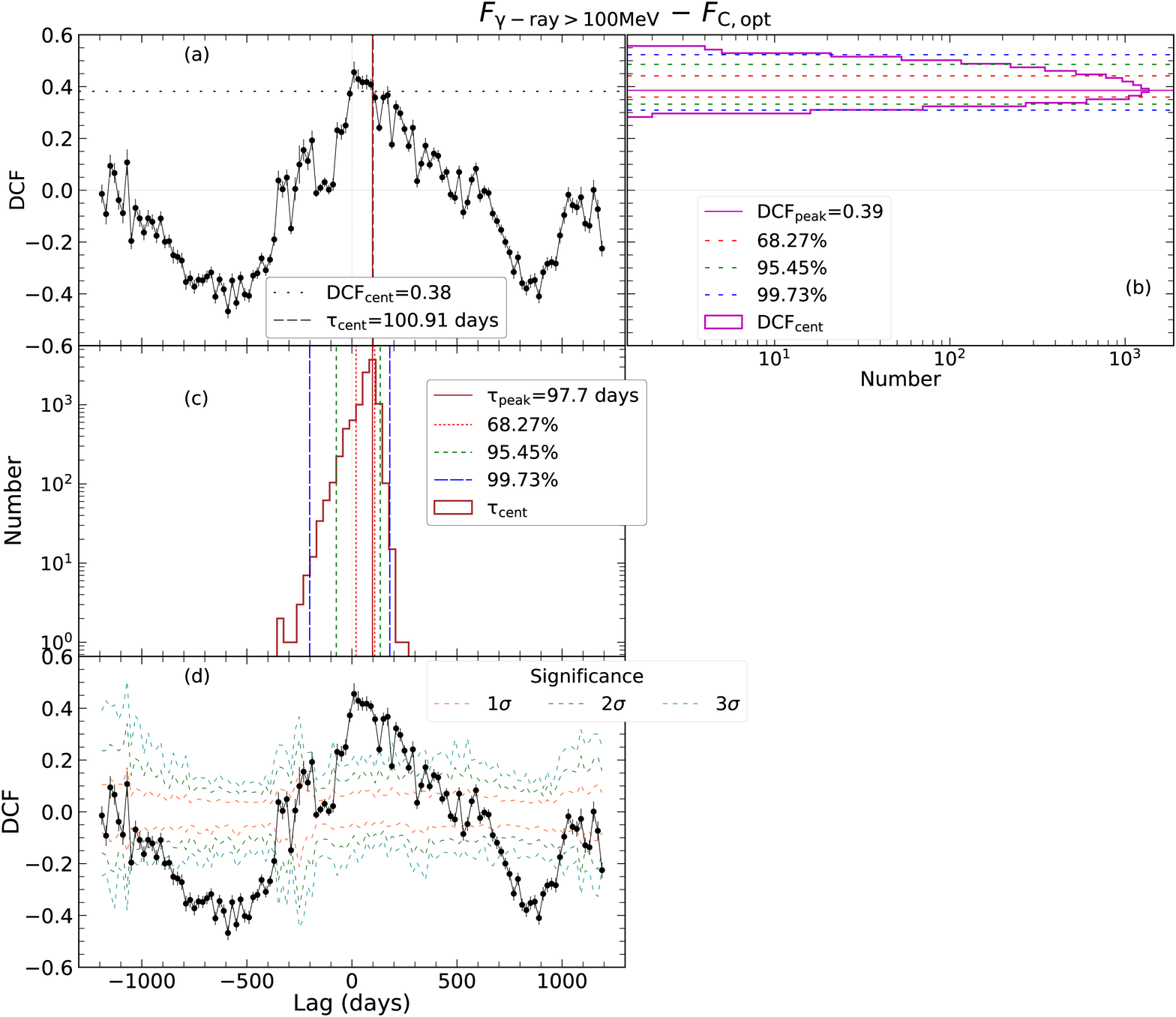}
     \caption{Results from bootstrapping method for $F_{\rm {\gamma}-ray>100~MeV}$ and $F_{\rm C,opt}$~(6024--6092~{\AA}) light curves. (a) PyDCF output with the centroid values of DCF (DCF$_{\rm cent}$; dotted horizontal line)  and lag $\tau_{\rm cent}$ (dashed vertical line) for the original data. (b) ${\rm DCF_{cent}}$ distribution obtained from cross-correlation results of the bootstrap output light curves. The peak of the distribution is represented by the  solid magenta horizontal line. The 68\%, 95\%, and 99\% confidence limits of ${\rm DCF_{peak}}$ are shown by red, green, and blue dashed lines, respectively. (c) Distribution of the centroid lag values obtained from cross-correlation analysis of the simulated light curves. The solid brown vertical line indicates the peak of the distribution ($\tau_{\rm peak}$). The confidence limits of $\tau_{\rm peak}$ are plotted in red, green, and blue dashed lines. (d) Plot showing the significance of the correlations at 1$\sigma$ (red dashed line), 2$\sigma$ (green dashed line), and 3$\sigma$ (blue dashed line) levels for different lags.}
     \label{fig:bootstrap_fermi_cont}
\end{figure*}

We determined the centroid lag from the average of the time lag values corresponding to the top 60\% of DCF values (for $\gamma-$ray$_{>100~MeV}-F_{\rm C,opt}$) mentioned above. The peak and confidence limits of the centroid time lags were also obtained in the same manner. The distribution of centroid lag values, peaking around 98 days, obtained from the bootstrap analysis is shown in Fig.~\ref{fig:bootstrap_fermi_cont}. The determined lag is, however, consistent with zero at 2$\sigma$ level. In addition, we tried the block bootstrap method to estimate the time-lag of correlation, considering that the data points in blazar light curves are mutually dependent. Block bootstrap is a simulation method used to estimate the distribution of test statistics \citep{lahiri2003resampling,gonccalves2011discussion}. Here we considered the non-overlapping block bootstrap method to re-sample time series data to estimate the correlation time-lag \citep{politis2003impact}. The original light curves were split into non-overlapping blocks of 10-day length. We then randomly re-sampled the data sets based on these blocks from the original data over 10,000 times to obtain the bootstrapped samples for each light curve. However, even with the block bootstrap method, we could not constrain the time-lag of correlation between the \textit{Fermi}-LAT and optical continuum light curves. The peak of the centroid lag distribution obtained through this method is also consistent with zero time lag at 1-$\sigma$ level.

Since \textit{Fermi} normally operates in an all-sky scanning mode, the $\gamma-$ray band light curve is obtained with a sampling time of 3 days. The optical data, however, suffer from observational gaps. There are two types of gaps 1) seasonal observing cycles and 2) gaps within the observing cycle. Therefore, we also estimated the correlations between the \textit{Fermi}-LAT $\gamma-$ray band and optical continuum by considering the optical observational cycles separately. However, we could not obtain a significant estimate of time lag for these seasonal light curves.

\section{Spectral energy distribution} \label{sec:sed}

PKS~1222+216 is known to show significant disk contribution in the optical/UV band \citep[e.g.,][]{2014MNRAS.442..131K}, apart from the synchrotron emission. To understand the variability of disk emission in the source, we generated optical--$\gamma$-ray SEDs for flaring (MJD: $\sim$56974.6--56977.8) and quiescent (MJD: $\sim$5651--55685) epochs in the $\gamma$-ray band. The $\gamma$-ray spectra were obtained from \textit{Fermi}-LAT observations, while the optical/UV and X-ray data were obtained from \textit{Swift} XRT and UVOT observations. The XRT spectra were generated using the online tool Build Swift-XRT products\footnote{\url{https://www.swift.ac.uk/user_objects/index.php}} \citep{2009MNRAS.397.1177E}. The corresponding UVOT filter data (dereddened) were obtained from the Multi-Mission Interactive service Archive provided by the Space Science Data Center (SSDC). The High energy Fermi-LAT spectra were generated as described in section \ref{sec:obs-Fermi}.

\begin{figure*}[htp]
     \centering
     \includegraphics[width=1.0\textwidth]{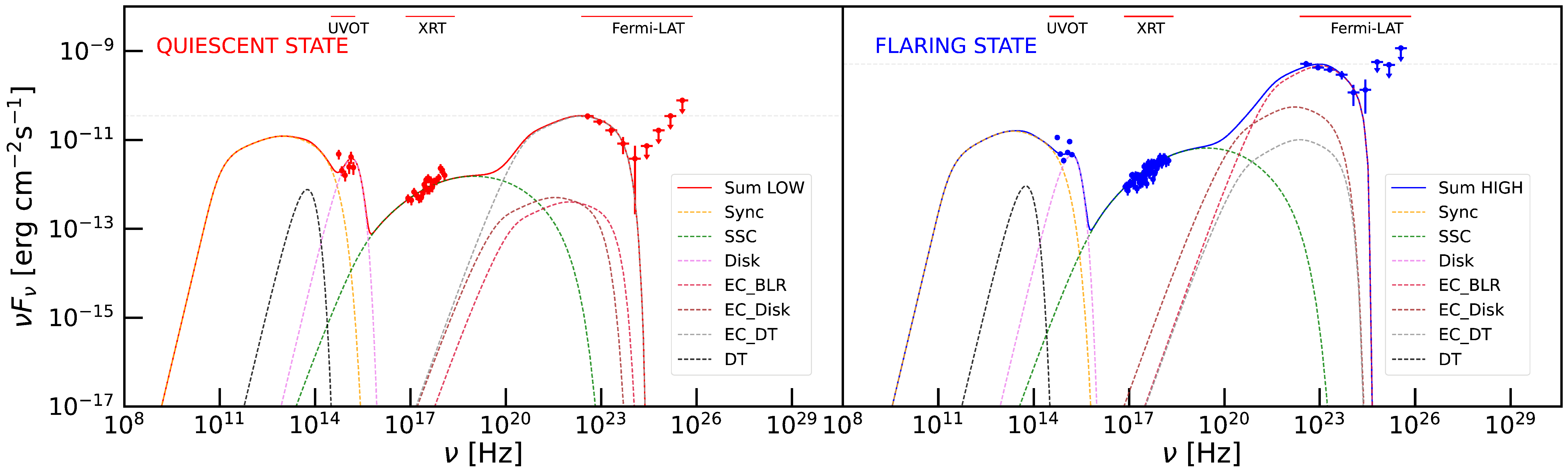}
     \caption{The SEDs of the source fitted with \textit{Jetset} code for the $\gamma$-ray flaring and quiescent epochs. The optical/UV photometric data were obtained from Swift UVOT filters (V, B, U UVW1, UVM2, and UVW2), and the X-ray (0.3--10~keV) spectra were taken from XRT observations. LAT spectra were used for the high energy $\gamma-$ray emission. The upper limits in LAT spectra are plotted as downward arrows.}
     \label{fig:sed}
\end{figure*}

The assembled data for the chosen epochs were used to investigate the spectral change during the flaring and quiescent states (see Fig.~\ref{fig:sed}). The double hump SED shape of blazars is usually interpreted within the leptonic scenario. Standard one zone leptonic scenario for FSRQ considers a single emission region covering the jet cross-section responsible for extended emission from IR to GeV energies. The variability period constraints the size of the emission region, usually within the BLR \citep{2009ApJ...692...32D,2009MNRAS.397..985G}; however, regions outside the BLR are equally viable \citep{2008ApJ...675...71S,2008Natur.452..966M}.We use the open source package \textit{Jetset} to model the broadband SED \citep{2020ascl.soft09001T,2011ApJ...739...66T,2009A&A...501..879T,2006A&A...448..861M}. Three components are considered to be contributing to the high energy emission in the second hump: disk, BLR, and torus. 

The disk emission is probably imprinted in the UVOT-data points. We use the highest frequency filter (UVW2) in the UVOT energy range to estimate the disk luminosity ($L_{disk}$) as in \cite{2011A&A...534A..86T}. This produces a lower limit of disk luminosity at $L_{disk} = 3.75 \times 10^{45}\,\rm{erg/s}$, which is of the same order of magnitude as in \cite{2011ApJ...733...19T}. The radiation from the disk photo-ionizes the BLR, modelled as a spherical shell of radius $R_{BLR}$ with inner and outer radii, $R_{BLR_{in}} \sim 0.9 R_{BLR}$ and $R_{BLR_{out}} \sim 1.1 R_{BLR}$, respectively. We set $R_{BLR}=10^{17}L_{d,45}^{1/2}$ as in \cite{2009MNRAS.397..985G}. A fraction of disk emission is re-emitted by the dusty torus. We consider torus to be emitting as a black body with temperature $T_{DT}=1000$~K approximated as a spherical volume of radius $R_{DT} = 7 \times 10^{18}$~cm.  We consider the emission region to be a sphere covering the entire cross-section of the jet and co-moving with the Bulk Lorentz factor. Considering a emission region outside BLR at $5.2 \times 10^{17}\,\rm{cm}$ corresponding to a variability time of 1~day for flaring epoch and $2 \times 10^{18}\,\rm{cm}$ for quiescent period corresponding to variability time of 4~days. The parameters used for the SED fitting are listed in table \ref{tab:fitting_parameter}. 
The value of N during SED fitting of flaring state increases by two orders of magnitude, hinting at a particle acceleration process (shock or magnetic reconnection). The dominating source of external seed photons for observed high energy emission seems to be the broad-line region during the flaring period. However, dusty torus contributed to the majority of observed high energy emission during the quiescent state.

\begin{table*}[]
\centering
\tabcolsep 0.3cm
\caption{Comparison of parameters from SED fitting on quiescent and flaring states.}
\label{tab:fitting_parameter}
\begin{tabular}{cccccc}
    \hline
    Parameter & parameter type & Quiescent & Flaring\\
     & & [MJD 55651 - 55685] & [MJD 56974 - 56978]\\
    \hline
    $R_{emm}$ & Size of emission region & $1 \times 10^{17}\,\rm{cm}$  & $2.6 \times 10^{16} \,\rm{cm}$\\
    $\Gamma_j$ & Bulk Lorentz factor & 10 & 10\\
    $\rm{\theta}$ & Jet viewing angle & 3 & 3\\
    z & Redshift & 0.432 & 0.432\\
    $L_{disk}$ & Disk luminosity & $3.75 \times 10^{45}\,\rm{erg/s}$ & $4.5 \times 10^{45}\,\rm{erg/s}$\\
    $T_{disk}$ & Peak disk temperature& $2.5 \times 10^{4}\,\rm{K}$ & $2.75 \times 10^{4}\,\rm{K}$\\
    $R_{BLR_{in}}$ & Inner radius of BLR & $1.9 \times 10^{17}\,\rm{cm}$ &  $1.9 \times 10^{17}\,\rm{cm}$\\
    $R_{BLR_{out}}$ & Outer radius of BLR & $2.3 \times 10^{17}\,\rm{cm}$ &  $2.3 \times 10^{17}\,\rm{cm}$\\
    $\tau_{BLR}$ & Fraction of disk luminosity reflected by the BLR & $0.15$ & $0.15$  \\
    $R_{DT}$ & Radius of the dusty torus & $7 \times 10^{18}\,\rm{cm}$  & $7 \times 10^{18}\,\rm{cm}$\\
    $T_{DT}$ & Dust temperature & $1000\,\rm{K}$ & $1000\, \rm{K}$\\
    $\tau_{DT}$ & Fraction of disk luminosity reflected by the orus &  $0.2$ & $0.2$\\
    $B$ & Magnetic field within emission region & $0.26\, G$ & $0.6\,\rm{G}$\\
    $N$ & Number density & 20 $/\rm{cm^3}$ & 220 $/\rm{cm^3}$ \\
    $\gamma_{min}$ & Low energy cutoff & 100 & $100$ \\
    $\gamma_{max}$ & High energy cutoff & $4 \times 10^{3}$ & $4.0\times 10^{3}$\\
    $\gamma_{break}$ & Turn over energy  & $800$ & $800$ \\
    p & Low energy spectral slope & 2.3 & 2.1\\
    $p_1$ & High energy spectral slope & 3.3 & 3.4\\
    \hline
\end{tabular}
\end{table*}

\section{Results} \label{sec:rslt}

\begin{table*}[]
\centering
\tabcolsep 0.3cm
\caption{$F_{\rm rms}$ for the parameters for different cycles and total period (Overall) of observations.}
\begin{tabular}{cccccc}
\toprule
\multirow{2}{*}{Cycle} & \multicolumn{5}{c}{$\rm F_{rms}$}       \\ \cmidrule{2-6}
 & $F_{H\gamma}$ & $F_{\rm C,opt}$~(6024--6092{\AA}) & $F_{\rm C,opt}$~(6380--6452{\AA}) & $\alpha$ & $F_{\rm \gamma-ray>100MeV}$ \\
 \midrule
C-3  &  0.08$\pm$0.01  &  0.3438$\pm$0.0003  &  0.348$\pm$0.001  &  0.21$\pm$0.01  &  0.51$\pm$0.03 \\
C-4  &  0.06$\pm$0.01  &  0.1429$\pm$0.0003  &  0.147$\pm$0.001  &  0.14$\pm$0.01  &  0.38$\pm$0.04 \\
C-5  &  0.05$\pm$0.01  &  0.0742$\pm$0.0003  &  0.080$\pm$0.001  &  0.11$\pm$0.01  &  0.49$\pm$0.04 \\
C-6  &  0.06$\pm$0.01  &  0.3235$\pm$0.0002  &  0.316$\pm$0.001  &  0.29$\pm$0.01  &  0.59$\pm$0.02 \\
C-7  &  0.07$\pm$0.01  &  0.1767$\pm$0.0002  &  0.175$\pm$0.001  &  0.16$\pm$0.01  &  0.63$\pm$0.04 \\
C-8  &  0.06$\pm$0.01  &  0.2185$\pm$0.0004  &  0.223$\pm$0.001  &  0.12$\pm$0.02  &  0.34$\pm$0.10 \\
C-9  &  0.02$\pm$0.01  &  0.0673$\pm$0.0003  &  0.066$\pm$0.001  &  0.15$\pm$0.02  &  -- \\
C-10  & 0.06$\pm$0.01  &  0.0232$\pm$0.0007  &  0.021$\pm$0.002  &  0.07$\pm$0.03  &  -- \\
\midrule
Overall & 0.084$\pm$0.004 & 0.4291$\pm$0.0001 & 0.4270$\pm$0.0003 & 0.235$\pm$0.004 & 0.92$\pm$0.01\\
(C-3 -- C-10) &  &  &  &  & \\
\bottomrule
\end{tabular}
\label{tab_frms}
\end{table*}

In this work, we studied the connection between the optical and $\gamma-$ray emissions from PKS~1222+216 monitored with Steward observatory and \textit{Fermi}-LAT. We have analyzed the long-term optical spectroscopic observations of the source. We estimated the broad H$\gamma$ emission line and the continuum fluxes for all the observations by fitting the spectra with \textit{powerlaw} and \textit{Gaussian} models in Sherpa. The optical continuum flux varies by a factor of $\sim$7 from about ${\rm 7.64\pm0.04\times10^{-14} erg~cm^{-2}s^{-1}}$ to ${\rm 52.39\pm\times10^{-13} erg~cm^{-2}s^{-1}}$. The power-law index $\alpha$ ranges from -0.32$\pm$0.07 to -1.56$\pm$0.13. The parameters of the Gaussian component range as follows: $F{_{\rm H\gamma}=(2.10\pm0.13~0-~3.69^{+0.14}_{-0.18}) \times 10^{-14} {\rm erg~cm^{-2} s^{-1}}}$, $FWHM{\rm = (2739.46^{+890.16}_{-161.48} - 5078.85^{+412.38}_{-406.15})~km~s^{-1}}$, $\lambda_{pos} = (6216.01^{+3.68}_{-0.84} - 6245.54^{+4.40}_{-4.25}$)~\AA, and \textit{EW} $= (2.96\pm0.08 - 31.42\pm1.56)$~\AA. We observed significant variability in the parameters that were confirmed by fitting the light curve of each parameter by a constant, which resulted in reduced $\chi^2 > 2$.  However, we note that the variability in the Gaussian parameters could also be contributed by the variable instrumental resolution of the spectra used in the analysis.

We measured the strength of variability of the parameters over the total monitoring period in terms of the fractional RMS variability amplitude ($F_{\rm rms}$) \citep{2003MNRAS.345.1271V}. The optical continuum emissions show significant variability with $F_{\rm rms}\sim0.4$, whereas the H$\gamma$ flux is less variable with $F_{\rm rms}\sim0.08$. The $F_{\rm rms}$ values of the parameters for the overall monitoring period and separately for different cycles are quoted in Table~\ref{tab_frms}. We calculated the equivalent width of the H$\gamma$ line in each observation from the continuum and H$\gamma$ emission line flux density for the spectral range. There is an inverse relationship between the equivalent width and the underlying continuum, as observed in other blazars \citep[e.g.][]{2018ApJ...866..102P}. The observed variation of \textit{EW} with the continuum flux is shown in Fig.~\ref{fig:opt_cont_line_ew_Fratio}. The inverse trend between these quantities suggests that the H$\gamma$ line is less variable than the continuum emission. Also, the flux density ratio between 5044${\rm \AA}$ and 6064${\rm \AA}$ decreases as the flux increases showing the "redder-when-brighter" behaviour (see Fig.~\ref{fig:opt_cont_line_ew_Fratio}). A similar trend has been previously reported for this object for the flux density ratio between 4700${\rm \AA}$ and 6600${\rm \AA}$ with V-band magnitude obtained from SO observations \citep{2011arXiv1110.6040S}. The other parameters obtained from the analysis did not show any direct correlations using the non-parametric Spearman's rank-order method.

The \textit{Fermi}-LAT light curve of PKS~1222+216, obtained roughly for the same observation period ($\sim2011-2018$), shows strong variability ($F_{\rm rms}\sim0.9$) with flaring events at various epochs. The optical continuum is also found to have a few flaring episodes. It is known that the optical and $\gamma-$ray emissions from blazars can be correlated with or without time-lag \citep[e.g.,][]{2014ApJ...797..137C, 2019MNRAS.490..124M}. To investigate the possible temporal correlation in optical and $\gamma-$ray bands, we performed the cross-correlation analysis of the $F_{\rm {\gamma}-ray>100~MeV}$/$F_{\rm C,opt}$~(6024--6092~{\AA}) light curves. The analysis provided no significant correlation for ${\rm H\gamma}$ emission with $\gamma-$ray continuum. The long-term optical continuum and LAT light curves show a moderately strong positive correlation, significant at 99\% level (see Fig.~\ref{fig:bootstrap_fermi_cont}). However, we caution that the time-lag of correlation is not significant at the two sigma level. Hence, we do not confirm a strong time-delayed correlation as the time-lag is too broad in this case. It is probable that the gaps between the observational cycles of optical monitoring dilute the possible strong correlations. We also carried out a similar analysis for the seasonal light curves, but no significant correlation was found among the light curves.

Previous studies \citep{2014ApJ...786..157A, 2014ApJ...797..137C} have analyzed the temporal correlation between the $\gamma-$ray and optical emissions from PKS~1222+216 at the flaring episode in 2010. \cite{2014ApJ...786..157A} found that $\gamma-$ray is leading the optical by $\sim$35 days with cross-correlation value of $\sim$ 0.4 whereas \cite{2014ApJ...797..137C} detected the $\gamma-$ray lead with a time-lag of $\sim$ 8.6~days. In a $\gamma-$ray (0.1–300 GeV), X-ray (0.2–10 keV), and optical (R band) cross-correlation study of a sample of low synchrotron peaked (LSP) and high synchrotron peaked (HSP) blazars, \citet{2019ApJ...877...39M} found multi-band variability with no time delay. Another work by \cite{2012MNRAS.421.1764S} studied the relation between the accretion rate and the jet power of a sample of blazars using SDSS and \textit{Fermi} observations. They found a clear positive correlation between the broad emission lines and $\gamma-$ray luminosities in Eddington units, supported by the Kendall test.

\section{Summary and Discussion} \label{sec:sum}

The high energy emission in FSRQs is expected to be produced by the external Comptonization scenario. The source of seed photons for EC process could be accretion disk, broad-line region or torus \citep[e.g.][]{1997ApJS..109..103D, 1999ApJ...515L..21B,2002A&A...386..415A}. Since the observed optical continuum luminosity is a combination of disk and jet emissions, the broad emission line luminosity in FSRQs provides an excellent probe of disk emission. In an attempt to study the relationship between various emission components in the FSRQ PKS~1222+216, we carried out a comprehensive investigation of the $\sim$8~year long optical/$\gamma-$ray monitoring observations, as described in the previous sections.

 We implemented a detailed spectral analysis to retrieve the optical emission properties that reveal strong variability in the optical continuum flux, whereas the H$\gamma$ line is less variable. The equivalent width of the H$\gamma$ line shows an inverse correlation with the optical continuum emission. This trend, along with the nearly constant line flux, show that the line emission is varying slowly compared to the underlying continuum as suggested by the well-known Baldwin effect \citep{1977ApJ...214..679B}. The source also exhibits a redder-when-brighter trend of decreasing flux density ratio ($F_{\rm 5044\AA}/F_{\rm 6064\AA}$) with an increase in the optical continuum flux. Since the synchrotron emission peaks at the infrared wavelengths in FSRQs, and the accretion disk contributes more to the bluer part of the SED, the redder-when-brighter behavior clearly points towards an  increasing contribution from the jet at high flux states in the optical band. Similar trend was observed by \cite{2011arXiv1110.6040S} in the same source and other blazars \citep[e.g.][]{2015RAA....15.1784Z}. All these trends indicate that the contribution of the accretion disk to the optical continuum variability is less significant in the source.
 
The variability studies of the various emission components clearly show that the $\gamma-$ray ($>{\rm100~MeV}$) flux is highly variable in the whole period of observation, with an $F_{\rm rms}$ of $\sim0.9$. The $\gamma-$ray emission appears to be more variable than the optical line and continuum components in the individual cycles of the optical monitoring program as well. The SED analysis and previous studies rule out the contribution of thermal emission disk emission to the variable $\gamma-$ray ($>{\rm100~MeV}$) emission. The higher variability in $F_{\rm{\gamma-ray>100MeV}}$ and the moderate positive correlation we observed between the $F_{\rm{\gamma-ray>100MeV}}$ and $F_{\rm C,opt}$~(6024--6092{\AA}) emissions can be attributed to the enhanced particle acceleration or cooling process at the flaring/quiescent epochs.

In this work, we have quantified the $\gamma-$ray/optical correlation in PKS~1222+216 using SPOL and \textit{Fermi}-LAT monitoring observations and studied the relationship between the emissions from the jet, disk, and BLR regions. The DCF analysis did not provide any strong correlation between H$\gamma$ line and $\gamma-$ray (${>100MeV}$) fluxes. As the broad emission lines arise due to the photoionization of the gaseous material by the disk photons in the BLR region \citep[e.g.][]{2019ApJ...876...49Z, 2011A&A...535A..73H, 2004ApJ...613..682P, 2000ApJ...533..631K, 1997ASSL..218...85N}, the aforementioned non-correlation most likely rules out the possibility of disk photons acting as the source of variability observed in $\gamma$-rays. This, in turn, implies that the variability of the $\gamma-$ray emission is most likely intrinsic to the high energy particles. The lack of strong variability of the H$\gamma$ line also suggests weak variability of disk component, and the observed correlation between $F_{\rm C,opt}$~(6024--6092{\AA}) and $F_{\rm{\gamma-ray>100MeV}}$ can be attributed to the jet contribution in both optical and $\gamma$-ray emissions. The observed correlation between $\gamma-$ray and optical continuum emissions is consistent with zero time-lag at the two sigma level. We note that the absence of a strong correlation and difficulty constraining the time-lags could be due to the gaps between the optical observing cycles. Obviously, studies with long-term high cadence monitoring can precisely determine the presence of time-delayed correlation in the source. 

The results from SED analysis indicates that the location of the emission region of size $2.5 \times 10^{16}\,\rm{cm}$ was at $0.2 \,\rm{pc}$ for the flaring epoch, whereas for the quiescent state an emission region of size $1\times 10^{17}\,\rm{cm}$ was apparently located at $0.6 \,\rm{pc}$. This suggests that the BLR photons significantly contribute to the high energy emission in the flaring epochs. An increased N during flaring epochs also hints at an increased number of high energy electrons through particle acceleration via shocks or magnetic reconnection. On the other hand, the dusty torus component apparently contributes to the seed photons for the EC process in the quiescent state. This can be explained by the variation in the energy density of the radiation fields (magnetic and external) as the distance from the centre increases. The observed results points to a scenario where the blazar output is dominated by the inverse Compton scattering of the external radiation, and the jet dissipation occurs at a distance of a few hundreds of Schwarzschild radius from the central black hole, as supported by the canonical jet model \citep{2009MNRAS.397..985G, 2009ApJ...704...38S}. Our result suggests that the $\gamma-$ray emission region lies in the BLR/torus regions in the FSRQ PKS~1222+216, in agreement with former studies on the source during its active and quiescent states \citep[e.g.,][]{2014ApJ...786..157A, 2021MNRAS.508.1986C, 2011A&A...534A..86T}. This is also consistent with some recent studies of other FSRQs \citep[e.g.,][]{2015ApJ...808L..48P, 2020NatCo..11.4176S}. In future work, we will explore the optical and $\gamma-$ray correlations of a large number of sources to establish the disk-jet connection in blazars.

\section{Acknowledgement} \label{sec:ack}

We thank the anonymous referee for the insightful comments. Data from the Steward Observatory spectropolarimetric monitoring project were used. This program is supported by \textit{Fermi} Guest Investigator grants NNX08AW56G, NNX09AU10G, NNX12AO93G, and NNX15AU81G. We acknowledge the use of \textit{Fermi}-LAT data and analysis tools from \textit{Fermi} Science Support Center. This work made use of data supplied by the UK Swift Science Data Centre at the University of Leicester. Part of this work is based on archival data, software or online services provided by the Space Science Data Center - ASI. SHE would like to thank Kavita Kumari for discussions on time-series analysis.


\bibliography{ref}{}
\bibliographystyle{aasjournal}



\end{document}